\documentclass[]{acm-sig-alternate}
\usepackage{lmodern}
\usepackage{amssymb,amsmath}
\usepackage{ifxetex,ifluatex}
\usepackage{fixltx2e} 
\IfFileExists{upquote.sty}{\usepackage{upquote}}{}
\ifnum 0\ifxetex 1\fi\ifluatex 1\fi=0 
  \usepackage[T1]{fontenc}
  \usepackage[utf8]{inputenc}
\else 
  \ifxetex
    \usepackage{mathspec}
    \usepackage{xltxtra,xunicode}
  \else
    \usepackage{fontspec}
  \fi
  \defaultfontfeatures{Mapping=tex-text,Scale=MatchLowercase}
  
\fi
\IfFileExists{microtype.sty}{\usepackage{microtype}}{}
\usepackage{color}
\usepackage{fancyvrb}

\DefineVerbatimEnvironment{Highlighting}{Verbatim}{commandchars=\\\{\}}
\newenvironment{Shaded}{}{}

\newcommand{\CommentTok}[1]{\textcolor[rgb]{0.38,0.63,0.69}{\textit{{#1}}}}

\newcommand{\NormalTok}[1]{{#1}}
\ifxetex
  \usepackage[setpagesize=false, 
              unicode=false, 
              xetex]{hyperref}
\else
  \usepackage[unicode=true]{hyperref}
\fi
\hypersetup{breaklinks=true,
            bookmarks=true,
            pdfauthor={},
            pdftitle={Distributed Versioned Object Storage -- Alternatives at the OSD layer},
            colorlinks=true,
            citecolor=blue,
            urlcolor=blue,
            linkcolor=magenta,
            pdfborder={0 0 0}}
\title{Distributed Versioned Object Storage -- Alternatives at the OSD layer}

\numberofauthors{3}

\author{
            \alignauthor Ivo Jimenez (Student) \\  \affaddr{UC Santa Cruz} \\   \email{ivo@cs.ucsc.edu} 
         \and
            \alignauthor Carlos Maltzahn \\  \affaddr{UC Santa Cruz} \\   \email{carlosm@ucsc.edu} 
         \and
            \alignauthor Jay Lofstead \\  \affaddr{Sandia National Laboratories} \\   \email{gflofst@sandia.gov} 
        }

\date{}

\makeatletter
\def\@copyrightspace{\relax}
\makeatother

\begin{document}


\maketitle


\begin{abstract}
The ability to store multiple versions of a data item is a powerful
primitive that has had a wide variety of uses: relational databases,
transactional memory, version control systems, to name a few. However,
each implementation uses a very particular form of versioning that is
customized to the domain in question and hidden away from the user. In
our going project, we are reviewing and analyzing multiple uses of
versioning in distinct domains, with the goal of identifying the basic
components required to provide a generic distributed multiversioning
object storage service, and define how these can be customized in order
to serve distinct needs. With this primitive, new services can leverage
multiversioning to ease development and provide specific consistency
guarantees that address particular use cases. This work presents early
results that quantify the trade-offs in implementing versioning at the
local storage layer.
\end{abstract}



\section{Introduction}\label{introduction}

Modern parallel and distributed storage systems encapsulate the storage
layer behind an object abstraction {[}1{]} since this allows to hide the
implementation details behind a key-valued interface: having
\texttt{get(oid)} and \texttt{set(oid, value)} functions where
\texttt{oid} is an object identifier. A device/service that exposes such
an interface is known as an Object Storage Device (OSD). In this work we
introduce the concept of a versioning OSD (VOSD): incorporating
versioning primitives as part of the OSD API. A VOSD stores multiple
versions of an object, allowing the user to execute time-travel
operations and to access an object's lineage.

The VOSD interface (shown below) requires a \texttt{version\_id}
parameter to be passed to any call. Additionally, similar
collection-wide operations can be implemented that allow handling
versions for a sets of objects (with a corresponding
\texttt{collection\_id} parameter). The regular OSD API can be supported
by assuming that \texttt{get/set} operations return/modify the latest
version.

\begin{Shaded}
\begin{Highlighting}[]
  \CommentTok{// object-level}
  \NormalTok{clone(v_id, o_id)}
  \NormalTok{get(v_id, o_id)}
  \NormalTok{set(v_id, o_id, value)}
  \NormalTok{diff(v_1, v_2, o_id)}
  \NormalTok{parent(v_1, o_id)}
  \NormalTok{children(v_1, o_id)}

  \CommentTok{// collection-wide}
  \NormalTok{clone(v_id, c_id) ...}
\end{Highlighting}
\end{Shaded}

A VOSD serves as a powerful building block, as distributed and parallel
storage systems can enable new services that leverage multiversioning,
allowing a user/application to choose from an spectrum of
multiversioning alternatives: distinct consistency needs can be served
depending on the use case. We look at two of these use cases next.

\section{Use Cases}\label{use-cases}

To exemplify the utility of having versioning as a first-class citizen
in an OSD interface, we look at two use cases: one distributed and
another one in a parallel setting. We focus on the issues at the single
OSD level since these are independent of scale.

\subsection{ACID Transactions}\label{acid-transactions}

In transactional database systems, versioning is usually employed to
implement optimistic concurrency control. In this setting, instead of
acquiring locks, every transaction operates over a snapshot of the
database in an isolated manner. When a transaction is ready to be
committed, a validation phase checks that it doesn't conflict with
others, in which case the transaction is aborted and has to be
restarted.

Implementing multiversion concurrency control (MVCC) {[}2{]} requires
keeping track of the highest-committed transaction (HCT). Access to this
record has to be serialized to avoid inconsistencies. Once this HCT
record is available (e.g.~as an object itself), implementing MVCC on top
of a VOSD is relatively straight forward. At the beginning of a
transaction, the collection of objects that is being transactionally
managed is snapshotted. At the end of the transaction, a lock is
acquired on the HCT record and every object in the isolated snapshot is
\texttt{diff}'ed against the corresponding HCT. If no conflicts arise,
the HCT is assigned to point to the new transaction and the lock is
released.

\subsection{Read-atomic Transactions}\label{read-atomic-transactions}

HPC applications use checkpointing as their main fault-tolerant
technique: periodically dump checkpoints to storage and, in the advent
of failures, recover by reading the latest checkpoint. A recent trend is
to provide asynchronous interfaces (e.g.~see the recent DOE FastForward
Storage and I/O effort {[}3{]}) to applications. Asynchrony allows an
application to request an I/O operation and not have to wait for its
completion. A challenge arises when multiple I/O operations depend on
each other, since in order to avoid inconsistencies (e.g. abort if a
dependant request fails), the user needs to keep track of these
dependencies and add new logic at the application level. All this extra
code introduces overhead and causes waste of computational resources.

By employing versioning, an HPC application can tag every I/O operation
with its corresponding checkpoint version and let multiple versions
co-exist. An out-of-core process can merge multiple versions or garbage
collect unused ones to free-up space. Additionally, similarly to the
MVCC case, a record that keeps track of the highest-readable checkpoint
(HRC) can be used to give analysis and visualization applications access
to consistent checkpoints (an isolation level known as read-atomicity
{[}4{]}).

\section{OSD Versioning Alternatives}\label{osd-versioning-alternatives}

There are mainly three alternatives for implementing an OSD API: by
using an in-memory backend, key-value store or a local POSIX filesystem.
Incorporating versioning to each of these can be done in distinct ways:

\begin{itemize}
\itemsep1pt\parskip0pt\parsep0pt
\item
  \textbf{POSIX}. If the underlying filesystem supports it,
  copy-on-write (CoW) can be used to represent multiple versions of an
  object. If filesystem lacks support for CoW, a VOSD can fall-back to
  having per-version copies.
\item
  \textbf{In-memory}. Copy-on-write memory can be employed. For complex
  objects this might carry an extra overhead. In such cases,
  alternatives like Ropes or Interning can be used.
\item
  \textbf{Key-value Store}. The most straight-forward way to implement
  it is by keeping a copy for each version of an object. This might be
  prohibitive for large objects.
\end{itemize}

We next present preliminary evaluation of implementations for each of
the above.

\section{Preliminary Results}\label{preliminary-results}

The Ceph distributed storage platform {[}5{]} provides an object
interface that exposes a \texttt{clone()} operation, allowing
applications to create snapshots of an object. Internally, Ceph
abstracts storage nodes as OSDs and currently supports the three backend
types mentioned earlier:

\begin{itemize}
\itemsep1pt\parskip0pt\parsep0pt
\item
  \textbf{POSIX}. A Ceph OSD can be backed by either XFS, ext4, ZFS or
  BTRFS. In our experiments we use XFS thus we cannot make use of a CoW
  operation.
\item
  \textbf{In-memory}. Ceph OSDs implement a custom in-memory store
  (MemStore), using CoW to back the snapshot operation.
\item
  \textbf{Key-value store}. The key-value store of a Ceph OSD is backed
  by an instance of LevelDB, which is what we use. Since leveldb doesn't
  support versioning, cloning an object results in making a full copy of
  an object.
\end{itemize}

\textbf{Experimental Setup}. Our experiments were conducted on a machine
with two 2.0GHz dual-core Opteron 2212, 8GB DDR-2 RAM, one 250GB Seagate
7200-RPM SATA hard drive, running Ubuntu 12.04. A Ceph OSD daemon runs
on the machine and a local client connects to it to operate on objects
stored in it. We measure two aspects: version creation and retrieval.

\subsection{Version Creation}\label{version-creation}

We generate a workload consisting of 100 objects and 100 versions. The
size of each object is 4MB. We modify a portion of the object for each
version (64 16KB chunks modified at random). We measure the time it
takes to create this workload for each backend. Table 1 shows the
results.

\begin{verbatim}
     Backend       Phase        Time (ms)
    ---------- --------------  -----------
     XFS            F               676
                    S                59
                    M             20829
    ---------- -------------  ------------
     MemStore       F               106
                    S                47
                    M              9247
    ---------- ---------------  ----------
     LevelDB        F               196
                    S               192
                    M              9548
\end{verbatim}

We break down the timings into three phases: \texttt{F} which
corresponds to the time it takes to create the first revision.
\texttt{S} denotes the average time that it takes to create a snapshot
of the collection. \texttt{M} corresponds to the average time it takes
to modify the 100 objects (64 16KB modifications for each object).

\subsection{Version Retrieval}\label{version-retrieval}

For the workload described above, we read the latest version of an
object, as well as a randomly selected version (in the {[}1,100{]}
range). The object being read is randomly chosen. We execute 100 queries
of each type and report the average. Table 2 shows the results.

\begin{verbatim}
     Backend     Latest (ms)   Random (ms)
    ----------- ------------- -------------
      XFS          11.5          11.7
      MemStore      3.2           3.1
      LevelDB       6.4           6.4
\end{verbatim}

\section{Future Work}\label{future-work}

As part of our ongoing project, we are defining a generalized
distributed multiversioning framework that will be able to support
multiple flavors of versioning. As mentioned previously, applications
can customize this service to their particular needs and observe
distinct consistency guarantees. We are currently looking at other use
cases that fit in this multi-versioned view: distributed softare
transactional memory, management of massive datasets, transactional
stream processing and programmable filesystems.

\section{References}\label{references}

{[}1{]} M. Mesnier, G. Ganger, and E. Riedel, ``Object-based storage,''
\emph{IEEE Communications Magazine}, vol. 41, Aug. 2003, pp. 84--90.

{[}2{]} P.A. Bernstein and N. Goodman, ``Multiversion concurrency
control---theory and algorithms,'' \emph{ACM Transactions on Database
Systems}, vol. 8, Dec. 1983, pp. 465--483.

{[}3{]} {FastForward} Team, ``FastForward storage and I/O design
documents.'' Feb. 2012.

{[}4{]} P. Bailis, A. Fekete, A. Ghodsi, J.M. Hellerstein, and I.
Stoica, ``Scalable atomic visibility with RAMP transactions,'' \emph{ACM
SIGMOD conference}, 2014.

{[}5{]} S.A. Weil, S.A. Brandt, E.L. Miller, D.D.E. Long, and C.
Maltzahn, ``Ceph: a scalable, high-performance distributed file
system,'' \emph{Proceedings of the 7th symposium on operating systems
design and implementation}, Berkeley, CA, USA: USENIX Association, 2006,
pp. 307--320.


\end{document}